# Long distance $\nu_e \to \nu_\mu$ transitions and CP-violation with high intensity beta-beams.

Carlo Rubbia

*GSSI/INFN, L'Aquila, Italy*

*Institute for Advanced Sustainability Studies, Potsdam, Germany*

## Abstract.

The recent experimental determinations [4-8] of a large $\theta_{13}$ phase have opened the way to a determination of the mass hierarchy and of the unknown CP-violating phase $\delta_{CP}$. Experiments based on horn produced neutrino and antineutrino conventional beams are presently under development [3][14]. The event rates are marginal for a definitive search, since they require very intense beams and extremely large detector masses.

Zucchelli [15] has proposed a method in which pure $\nu_e$, $\bar{\nu}_e$ beams are generated by the $\beta$-decay of relativistic radio-nuclides stored in a high energy storage ring pointing towards a far away neutrino detector. Since they have a much smaller beam transverse momentum distribution, the neutrino flux will be much more narrowly concentrated than with a horn. The isomeric doublet Li-8 ($\bar{\nu}_e, \tau_{1/2}$ = 0.84 s) and B-8 ($\nu_e, \tau_{1/2}$ = 0.77 s) has been studied. Neutrino and antineutrino beams are produced with the small average transverse momentum of about 6.5 MeV/c.

Radioactive ions may be generated with the help of a dedicated table-top storage ring in order to supply a suitable ion source to be accelerated at high energies either at FNAL or at CERN. Ions should then extracted from the accelerator and accumulated in a decay storage ring with a long straight section pointing toward the neutrino detector.

A massive detector based on the liquid Argon technology [1] is probably offering the best opportunities for such future programme. The present ICARUS LAr-TPC experiment [1] has already collected at the LNGS a number of events in the relevant neutrino energy region. They should provide [22] a first evidence for a conclusive experimental study of the competing signals and more generally for the actual feasibility of the $\beta$-beam option in a search of the CP violating phase $\delta_{CP}$. Additional data may be provided in the near future with the ICARUS and MicroBooNe neutrino experiments located at a short distance neutrino beam and that will collect a much larger number of neutrino (CC and NC) events.

*June 6th, 2013*

# 1.— Introduction.

As well known, crucial but very difficult experiments are needed in order to complete the phenomenology of the neutrino sector. Such a wide programme demands new accelerated neutrino beams with well identified initial species and a long distance between beam and detector. A massive detector based on the liquid Argon technology [1] is probably offering the best opportunities for such future programmes. Several other proposals have more conservatively chosen a very massive water Cherenkov detector [2] or a finely segmented liquid scintillator [3].

Cosmological arguments have suggested that in order to build up the today's dominance of matter over anti-matter in non-equilibrium conditions the CP violation in the quark sector should be extended to the lepton sector. To this effect, all the three neutrino mixing angles must have non zero values, including the recently observed $\theta_{13}$, where results have been reported by T2K [4], MINOS [5], DOUBLE CHOOZ [6], RENO [7] and especially DAYA-BAY [8]. The different experiments are in good agreement and the most precise is the one of DAYA-BAY with the result $\sin^2(2\theta_{13}) = 0.0890 \pm 0.0112$, equivalent to $\sin^2(\theta_{13}) = 0.023 \pm 0.003$ or $\theta_{13} = (8.7 \pm 0.6)°$.

Consequently with a sufficient statistics and in absence of competing backgrounds the lepton CP violating phase $\delta_{CP}$ may become more easily accessible. Starting either from an initial high purity either $\nu_\mu$ or $\nu_e$ source, the key physical process is the observation of the tiny $\theta_{13}$ related oscillation mixing between $\nu_\mu \leftrightarrow \nu_e$ around a energy dependent distance L corresponding to an approximate neutrino flight path of about 2.0 MeV/m  The oscillation mixing probability, in absence of matter effects and $\theta_{13} = 9°$ is shown in Figure 1 for different values of the CP violating phase as a function of the reduced distance in MeV/m.

The relevant energy domain is therefore directly proportional to the distance between the source and the decay locations, independently of the method chosen to produce the neutrino.  The values of the maximum of the first peak of the oscillation are $1470 \pm 295$ MeV for the LNGS [9] and Soudan [10] neutrino laboratories (L = 730 km), $434 \pm 120$ MeV for the SK [11] experiment (L = 295 km), $260 \pm 55$ MeV for the perspective Frejus [12] laboratory (L = 130 km) and $2600 \pm 520$ MeV for the FNAL to Homestake (L = 1290 km) configuration [13].

For too short L configurations, the neutrino energy spectrum is strongly affected by nuclear corrections. For too long L configurations, matter oscillation effects and eventually direct $\nu_\tau$ production become relevant and alter the $\theta_{13}$ oscillation pattern, introducing large differences between initial $\nu$ and $\bar{\nu}$ depending of the knowledge of the actual matter density underground. Distances of the order of about $1000 \pm 300$ km represent an acceptable compromise between decay distance, effects due to matter oscillations and absence of direct $\nu_\tau$ production. Therefore both LNGS/CERN and FNAL/Soudan (L = 730 km) or FNAL/Homestake (L= 1290 km) represent optimal choices.

The massive neutrino detector is located at a fixed distance. In order to cover completely the $\theta_{13}$ oscillation pattern, a wide neutrino energy spectrum must be observed. The sizeable event rate persists over successive maxima with n = 1, 2, 3, and so on. Indeed the most significant differences in CP related values occur for n > 1. Therefore the full accessibility to an extended neutrino energy region is highly relevant for a sensitive search of CP-violation. The importance of the secondary maxima with n > 1 is clearly evident.

## 2.—    Conventional beams.

Observations of $\nu_\mu \rightarrow \nu_e$ oscillations of a beam composed initially of muon neutrinos over a long baseline and a wide range of neutrino energies can be used in order to determine the mass hierarchy (the sign of $\Delta m_{32}^2$), and the unknown CP-violating phase $\delta_{CP}$. Several next generation experiments based on the horn produced neutrino and antineutrino beams at different distances from the target are presently under active development, based on the combination of an adequate baseline, a very large detector mass and a wide-band beam and energies. Such precision measurements of the 3-flavor neutrino oscillation parameters are more generally addressed to the search for new physics that may manifest itself as deviations from the expected neutrino model.

The LBNE [14] planned experiment at L=1290 km in the Homestake with the beam from FNAL is taken as a reference. Figures 2a and 2b show the expected spectra of $\nu_\mu \rightarrow \nu_e$ and $\bar{\nu}_\mu \rightarrow \bar{\nu}_e$ oscillation events in an initial 10-kton LArTPC for 5 years of neutrino and 5 years of antineutrino running with a 700 kW incoming proton beam for $\sin^2(2\theta_{13})= 0.1$ and assuming normal mass ordering. Backgrounds from intrinsic beam $\nu_e$ (cyan), $\nu_\mu$-NC (yellow), and $\nu_\mu$ CC (green) are displayed as stacked histograms. The expected number of events are relatively small, $N_{\nu e}^{signal}$=180 and $N_{\bar{\nu}e}^{signal}$ = 90 for $\delta_{CP}$ = 0. The red (blue) histogram are the $N_{\nu e}^{signal}$=220 (131) and $N_{\bar{\nu}e}^{signal}$ =68 (99) with $\delta_{CP}$ = -90° (90°) respectively. In spite of the very long duration of the exposures, event rates are rather small. The number of expected signal events with $E_\nu \leq 1.5$ GeV is statistically insignificant and dominated by backgrounds. The significance of a signal is therefore limited to the observation in the region with $E_\nu \geq 1.5$ GeV namely only around the first maximum.

The accuracy of the experimental result may be improved increasing the LArTPC fiducial mass and/or the proton beam power. In Figure 3 the significance (σ) of $\delta_{CP}$ has been given for masses of 10, 15 and 34 kton. Under any circumstances, the ability of determining $\delta_{CP}$ with an appropriate statistical evidence (ultimately ≥ 5 σ) strongly depends of the— so far unknown— actual value of $\delta_{CP}$.

### 3.—    The Zucchelli's initial beta beam proposal.

Zucchelli [15] has proposed an interesting method in which pure $\nu_e$, $\bar{\nu}_e$ beams are generated by the $\beta$-decay of relativistic radio-nuclides stored in a high energy storage ring pointing towards the neutrino detector. In contrast with a conventional beam, the advantage of this method is that very pure $\nu_e$, $\bar{\nu}_e$ beams are produced with a nearly zero initial contamination. The main background source $O\left(\approx 10^{-5}\right)$ is caused by neutrino produced by nuclear interactions of the decay daughter nuclei ejected from the beam and interacting with the walls of the storage ring.

The angular spread of the accelerated and stored relativistic ion beam is very small when compared to the one caused by the radioactive decay angle. Hence the analysis of the beam is very simple and neutrino fluxes at the detector can be easily determined with the help of the kinematics of the radioactive decay. There is no a priori need for a near-detector. With the radioactive ion beam pointing to the detector, the neutrino energy $E_\nu$ has a sharp end point given by $E_\nu = 2\gamma Q^*$, where $Q^*$ is the CM neutrino end-point energy and $\gamma$ is the relativistic factor of the stored ions. The neutrino energy spectrum at the detector is therefore simply $2\gamma$ times the neutrino spectrum of the decaying radio-nuclide in the centre of mass. The opening angle of the neutrino is $\approx 1/\gamma$, independent of $Q^*$.

One of the main advantages of the $\beta$-decay driven beam is the much smaller beam transverse momentum distribution when compared with the one of the conventional horn driven target. Therefore the neutrino flux and the rate of recorded events may be much more favourable than the one in the case of a horn. The main requirement is the availability of both $\nu_e$ and $\bar{\nu}_e$ beams and of a pair of low mass radio-nuclides with an adequately short half-life — in the order of 1 s — and which can be produced in an sufficiently large quantity in order to ensure the required event rate at many hundreds of kilometres away.

The original Zucchelli proposal [15] has been based on anti-neutrino from He-6 ($Q^* \approx 3.508$ MeV and $\tau_{1/2} = 0.806$ s) and on neutrino from Ne-18 ($Q^* \approx 4.443$ MeV, $\tau_{1/2} = 1.67$ s). According to the CERN proposal, He-6 is generated by a high current superconducting proton LINAC on a neutron spallation target in the several MWatt range. The rate for which Ne-18 has been chosen is more uncertain and most likely the resulting event rate may be substantially (as much as one order of magnitude) smaller.

The He-6 beam [16] may be accelerated in the SPS to an equivalent proton energy of $\approx 300$ GeV and accumulated in an additional storage ring with two very long straight sections in which ions spontaneously decay with about 80 s half-life. The relativistic factor is $\gamma = Z/A\gamma_{proton} = \gamma_{proton}/3 \approx 100$. The end point energy of anti-neutrino in the laboratory is therefore 700 MeV.

A massive Cherenkov water detector may be located for instance inside the Frejus tunnel (L = 130 km), where the first $\bar{\nu}_e \rightarrow \bar{\nu}_\mu$ oscillation maximum occurs at 260 MeV. According to the CERN design parameters, the number of collected He-6 ions in the storage ring may be of the order of $10^{11}$ ions/s (3 x $10^{18}$ ion/year), corresponding to about 70 CC un-oscillated $\bar{\nu}_e$/kton/y. Therefore in order to collect a total of about $10^5$ ev/year, the water fiducial mass should be near to one Megaton.

The choice of such a short decay path (L = 130 km) are not without problems. For neutrino energies not too far from threshold, the elastic cross section for Oxygen is strongly suppressed because of the Pauli factor that requires that the outgoing nucleon should be above the Fermi sea. The fast energy dependence of the cross sections near threshold requires a very accurate energy resolution and seriously complicates the inter-calibrations for the $\mu$ and $e$ related channels. Individual cross sections are not sufficiently well known and they are presently only roughly estimated theoretically with the help of nuclear physics arguments. They must be experimentally measured with extensive calibration experiments. It is debatable if the energy dependence of neutrino oscillation pattern could be ultimately determined with a sufficient reliability very near to the threshold.

The muon-oscillated low energy events not well separated from the large background of muon events due to atmospheric neutrinos. In order to remove such a background, the beam stored in the ring must be very tightly bunched, with a duty cycle factor that in the CERN design exceeds a factor of 1/1000. This introduces additional limitations to the parameters of the storage ring and to the preceding chain of the many successive particle accelerators.

The source initially producing $\nu_e$ (instead of $\nu_\mu$) introduces a large background due to the presence of neutral currents events which may generate a $\pi^+$ for instance via the reaction $\bar{\nu}_e + p \rightarrow \bar{\nu}_e + \Delta^+ \rightarrow \bar{\nu}_e + n + \pi^+$, competing with the smaller rate of oscillated $\mu^+$ events, $\bar{\nu}_e + p \rightarrow \mu^+ + n$. In a water Cherenkov detector of such a huge mass low energy $\pi^+$ and $\mu^+$ particles will have nearly identical signatures, At $\langle E_\nu \rangle \approx$ 350 MeV kinematic separation is not sufficient and the required background rejection just with the help of kinematics cuts is, in our view, unproven at this stage.

## 4.—    Beta beams with the isomeric doublet Li-8 and B-8.

The aim of the present paper is the one of choosing other radioactive nuclei with the highest possible CM decay energies and a half-life comparable to the one of He-6. A new method of accumulating an adequate number of radioactive nuclei has been recently described [17]. In particular the case of the isomeric doublet Li-8 ($\bar{\nu}_e$, $\tau_{1/2}$ = 0.84 s) and B-8 ($\nu_e$, $\tau_{1/2}$ = 0.77 s) has been examined, (Figure 4). Two body production reactions are Li-7(d,p) Li-8 and Li-6(He3,n) B-8. In the region of few MeV the typical cross section for the reaction Li-7(d,p) Li-8 is about 100 mb with (a maximum at 200 mb) while it is about a fraction 10 lower for the reaction Li-6(He3,n) B-8.

The B-8 beta decay spectrum $B_5^8 \rightarrow Be_4^8 + \beta^+ + \nu_e$ is well studied since it is the dominant high energy Solar Neutrino spectrum. It has a complicated decay scheme that has been described by Bahcall *et al*. [18]. The 2+ state of B-8 decays $\beta^+$ in 96 % of the times with 17.979 MeV to a very broad 2+ state at 3.04 MeV, which in turn decays into $2\alpha$. The neutrino energy distribution is somewhat different than one of a simple allowed $\beta$-spectrum with an end point around 15 MeV and it has been numerically calculated in Ref. [18]. In analogy with the Zucchelli's proposal on He-6, $\bar{\nu}_e$ may also be produced with Li-7(d,p) Li-8 ($\tau_{1/2}$ = 0.84 s). Li-8, decaying to the same Be-8 final state, is identical to B-8 at least in the strong interaction limit (no Coulomb corrections). The observed neutrino distribution is therefore very similar and it can be determined experimentally.

The described ionization "cooling" [17] is intended to produce a large number of slow ($v \approx 0.1c$) secondary ions with the help of a tiny "table top" storage ring (Figure 5). Incoming singly ionized Li-7 or Li-6 ions are fully stripped and permanently captured in the ring with a thin gas jet target. The energy loss in the gas jet is compensated by an accelerating RF cavity. "Cooled" particles stably circulate in the beam and produce Li-7(d,p) Li-8 or Li-6(He3,n) B-8 interacting again with the same thin gas foil. It is proposed to produce reactions in the "mirror" kinematical frame, namely with the heavier ion colliding against the gas target. Kinematics is very favourable since secondary ions are emitted in a narrow angular cone (around $\approx$ 10 degrees for the above reactions) and with a low energy spectrum.

Ionization cooling [17] is achieved in steady conditions and many traversals through the thin foil in equilibrium between ionisation losses, typically of few hundred $\mu g/cm^2$ and the gain of a RF-cavity. However with a uniform target "foil", while transverse betatron oscillations are "cooled", the longitudinal momentum spread diverges exponentially since faster (slower) particles ionise less (more) than the average. In order to "cool" also longitudinally, a "chromaticity" has to be introduced with a radially shaped wedge "foil", such as to increase (decrease) the ionisation losses for faster (slower) particles. Multiple scattering and straggling may be then effectively "cooled" in all three dimensions with a method similar to the one of synchrotron cooling but adequate for low energy ions [17].

Particles therefore stably circulate in the beam until they may undergo some nuclear processes in the thin target foil. The range penetration of the secondary ion is very short, typically some tens of micron of solid material. Initial ions are captured as neutral atoms in a thin layer of the appropriate material. At a sufficiently high temperature and with an appropriate choice of the foil, atoms are spontaneously liberated in a time period which is short compared to the ion decay lifetime (a fraction of a second). They then become the ion-source for the subsequent acceleration of the beta-beam. The technique of using very thin targets in order to produce and to quickly extract at very high temperatures ($\approx$ 2000 K) the secondary neutral beam has been in use for many years. Probably the best known and most successful such a source of radioactive beams is ISOLDE [19].

A critical point is the delay of the collected ions to the catcher-ion-source system (CISS). This requires the understanding of the delay causing processes. The CISS is made of a small ring shaped thin box at about 10 degrees around the primary beam direction with a hole for the circulating ion beam and with a number of thin catcher foils inside (Figure 5). The release of the nuclide proceeds in two subsequent steps: *diffusion* from the place of implantation to the surface of the solid state catcher and *effusion* in the enclosure until the emission as a neutral particle. Solid state *diffusion* is governed by Fick's law. For an initially homogeneous distribution with diffusion coefficient $D$ in a thin foil of thickness $d$, the release efficiency for an ion of half-life $\tau_{1/2}$ is

$$Y(\tau_{1/2}) = \tanh\left(\sqrt{\lambda \pi^2 / 4\mu_o}\right) \Big/ \sqrt{\lambda \pi^2 / 4\mu_o} \approx 0.76\sqrt{\mu_o \tau_{1/2}} \ ,$$

with $\mu_o = \pi^2 D / d^2$ and $\lambda = \ln 2 / \tau_{1/2}$. Experimental information on the value of $D$ for Lithium ($\bar{v}_e$) in various (hot) materials with an appropriate CISS are shown in Figure 6. It is believed that at sufficiently high temperatures ($\approx 2000$ °C) an acceptable value ($\tau_{1/2} \leq 0.5$ s) with $10^{-12} < D(m^2/s) < 10^{-10}$ is possible. The situation for Boron is unknown at present and R&D is needed to assess its feasibility.

With Li-6 or Li-7 nuclei and at the energies of interest, the nuclear elastic and inelastic reaction cross sections producing the ejection of the particle from the beam are typically of the order $\sigma_{loss} \approx 10^{-24}$ $cm^2$. The corresponding integrated target thickness for 1/e absorption in $D_2$ and He-3 are respectively $\approx 3.3$ and $\approx 5$ g/cm$^2$, to be compared with a typical foil thickness of a fraction of mg/cm$^2$. With a jet target thickness of the order of 0.33 mg/cm$^2$, the nuclear 1/e lifetime is of about n $\approx 10^4$ turns, namely the average duration of a circulating particle inside the DC ring is of the order of 1 ms, with the incoming beam progressively captured by the nuclear process.

The typical cross section for the specific reaction Li-7(d,p) Li-8 is about 100 mb, while for the reaction Li-6(He3,n) B-8 it is about a fraction 10 lower. Therefore the useful production yield is respectively about 10% and 1% of the ejection Li beam yield. The resulting accumulation rate is between three and four orders of magnitude greater than the one of a conventional thick target configuration and a single beam passage.

The resulting activity is due mainly $\alpha$ and $\beta$ with neutrons for the (not precisely known) alternate reaction Li-7(d,n) Be-8, which decays instantly within the gas jet into 2 $\alpha$ and Li-7(d,2n) Be-7 with a secondary cross section of about 100 mb. Be-7 has a half-life of 53 days and it decays $\beta^+$ back to Li-7 with a $\gamma$ line at 477 keV.

In practical unit and for an order of magnitude estimate, we may assume for the reaction Li-7(d,p) Li-8 a raw rate of $10^{14}$ events/s, corresponding to a beam injection in the storage ring of about $10^{15}$ ion/s. The initially injected current of singly ionised particles before the stripping is therefore $\approx 160$ µA. Since the incoming beam lifetime is $\approx 1$ ms, The average current circulating in the ring is about a factor $10^{-3}$ smaller, corresponding to $10^{12}$ ions for Li-7(d,p) Li-8, namely a circulating ion current (Z = 3) of 3.5 Ampere with a revolution frequency $f_S = 7.35$ Mhz. The required power of the injected beam is relatively small, $\approx 4$ kW for T = 25 MeV. However the power due to the ionisation losses in the gas jet, to be compensated by the re-circulating RF with an energy loss of 300 keV/turn is much larger, 1.06 MWatt. The gas jet must therefore dissipate such a large amount of power.

The reaction Li-6(He3,n) B-8 is far less known and it is likely to be limited to a much smaller intensity essentially because of (1) the smaller production cross section and (2) the availability of the necessary amounts of He-3 gas. This may be partially compensated by the larger neutrino vs. antineutrino cross sections.

At this stage these figures should be considered as purely indicative, since the realization of a practical accumulator will strongly depend from the success of an extensive future R&D.

## 5.— An accelerator/storage setup based on β-decays of B-8 and Li-8.

The B-8 (Z = 5, A = 8) decays in 96% of the events with an end point at Q = 17.79 MeV and an average electron energy $\langle e^+ \rangle$ = 6.55 MeV. The corresponding decay of Li-8 (Z = 3, A = 8) is at Q =16 MeV and $\langle e^- \rangle$ = 6.24 MeV. In order to perform a sensitive search for CP violation, the experimental observation of the $\bar{\nu}_e \rightarrow \bar{\nu}_\mu$ and $\nu_e \rightarrow \nu_\mu$ reactions must be extended well around the region of the main maximum of Figure 1 and should include at least the minimum and the second peak, the sensitive region to $\delta_{CP}$. This is made in principle possible by the extreme cleanliness of the beta decay source. Therefore the appropriate distance L of the detector from the neutrino source and the corresponding neutrino energy spectrum must be provided. Two indicative L values have been considered, namely CERN/ LNGS and FNAL/Soudan (L = 730 km) and FNAL/Homestake (L = 1290 km). (Figures 7a and 7b).

The main indicative parameters of the decay storage ring (SR) are given in Table 1. The transfer from the accelerator to the SR may be performed with a method similar to the one of one of the rings of the CERN-ISR [20] and of the RHIC heavy ion storage rings at the Brookhaven National Laboratory [21]. The ISR stored proton currents in physics runs were typically 30 to 40 A (20 A design values) with a maximum achieved stored current of 57 A. The SR circulating currents of Table 1 (17 to 47 A) are therefore similar to the one of the protons of the ISR but with a much larger and superconducting ring. Beams of this intensity are achieved stacking many successive pulses next to one another (see Figure 8). For this purpose a radiofrequency system is needed. After a pulse has been injected the RF system accelerates the protons just enough to move the particles from their injection orbit to an orbit nearer to the outside of the vacuum chamber. When the acceleration has been done, the injection orbit is free to receive the next pulse width, which in turn is accelerated and moved to an orbit only a fraction of mm from where the previous pulse was left. This stacking process can be repeated indefinitely in the ring. The unstable ions progressively decay with a relativistic lifetime which varies between some 60 and 120 s depending on the configuration. The surviving fraction of ions are removed from the far edge of the aperture available in the vacuum chamber with a presumably internal beam dump oriented toward the direction opposite from the one of the decay of the neutrino beam.

Both the FNAL and CERN existing accelerator complexes have been considered. Fully stripped Li-8 and B-8 beams are extracted from the accelerator and accumulated on a storage ring with one long straight section pointing to the neutrino detector, comprehending 1/3 of its circumference. The relativistic factors of ions are $\gamma_{B-8} = Z/A \gamma_{proton} = 5/8 \gamma_{proton} = 0.625 \gamma_{proton}$ and $\gamma_{Li-8} = 3/8 \gamma_{proton} = 0.375 \gamma_{proton}$. The event $N_{\nu e \to \nu \mu}$ and $N_{\bar{\nu} e \to \bar{\nu} \mu}$ rates have been estimated for an indicative accelerated ion intensity of $10^{13}$ ion/s and a far away LAr-TPC detector of 34 kton. Data are integrated over a total of 2x1000 effective days. (5 y at 200 ev/y each for neutrino and anti-neutrino).

The Main Energy Injector at FNAL with the ion equivalent to 120 GeV protons corresponds to a relativistic factor $\gamma_{B-8}(\nu_e) = 80$ for B-8 and the smaller factor $\gamma_{Li-8}(\bar{\nu}_e) = 48$ for Li-8 corresponding to the neutrino average energies $\langle p_e \rangle = 1047$ MeV and $\langle p_e \rangle = 628$ MeV. Therefore the resulting neutrino spectra are generally insufficient for a comprehensive search of CP violating phase $\delta_{CP}$ and only B-8 at L = 730 km may offer an acceptable number of oscillated $\nu_e \to \nu_\mu$ events, $N_{\nu e \to \nu \mu} = 1089$ with $\delta_{CP} = 0$. The muon oscillated spectrum is shown in Figure 9 for both values of L.

The CERN accelerator complex offers a much better choice because of its higher available energies. Two equivalent proton energies have been considered: for B-8 we have chosen $E_p = 190$ GeV corresponding to $\gamma_{B-8}(\nu_e) = 126$ and $\langle p_e \rangle = 1658$ MeV ; for Li- 8 we have chosen $E_p = 400$ GeV corresponding to $\gamma_{Li-8}(\bar{\nu}_e) = 160$ and $\langle p_e \rangle = 2094$ MeV, all well matched to the expectations of Figures 10a and 10b.

Some of the main parameters of the accelerator, the storage ring and the detector are listed in Table 1 both for FNAL and CERN. The accelerator complex is fundamentally unchanged, with however the main difference in the production of the radioactive ion source. A significant amount of radiation is produced by the ions during acceleration. The main decay channels are a broad $\alpha$ peak and the $\beta^+$ decays with some associated internal bremsstrahlung $\langle IB \rangle \approx 40 keV$ (in the ion rest frame). In addition and in analogy with the ordinary acceleration some nuclear interactions occur with the walls of the accelerators.

The total neutrino events collected from the presently described beta-beam are substantial for the above indicated parameters, typically of the order of 1 to 3 events/hour, depending on the distance and beam configurations (see Table 1).

The muon oscillated spectra are shown in Figures 10a and 10b for the two above chosen distances and $10^{13}$ ion/s. The numbers of recorded $\nu_e \to \nu_\mu$ events integrated over a total of 1000 effective days are $N_{\nu e \to \nu \mu} = 4225$ (2109) and $N_{\bar{\nu} e \to \bar{\nu} \mu} = 1448$ (546.6) for L=730 km (L=1290 km). In absence of sizeable backgrounds, the observations will therefore cover the full region relevant to the CP violating phase $\delta_{CP}$ with an adequate statistics and with a reasonably sized detector.

### 6.— The properties of the events in a LAr-TPC.

The $\nu_e \rightarrow \nu_\mu$ process production in the elastic and inelastic CC channels for both neutrino and anti-neutrino events will be observed in a large LAr-TPC detector. The main signature relevant to the CP violating phase $\delta_{CP}$ is coming from $\nu_e \rightarrow \mu^- X$ from B-8 and $\bar{\nu}_e \rightarrow \mu^+ X$ from the Li-8 initiated beam.

The main competing background is the production of charged pions which may emulate the presence of the muons. The main processes are the neutral current produced inelastic processes leading to a negative pion capture or positive pion decay, with the pion misidentified into a decaying or eventually captured muon. (Figure 11) These neutral currents backgrounds [23] have an experimentally observable rate of the order of 1/10 of the CC current signal without oscillations, although with substantial uncertainties due to the poor experimental spectral information in the literature. Therefore NUANCE simulations [24] have been conveniently used.

With the now observed value of $\sin^2(2\theta_{13}) = 0.09$ and the distributions of Figures 11a and 11b, the average fractions of the electron to muon oscillated signals are $(\nu \rightarrow \mu^-)/(\nu \rightarrow e^-) \approx 1/20$ (B-8) and $(\bar{\nu} \rightarrow \mu^+)/(\bar{\nu} \rightarrow e^+) \approx 1/40$ (Li-8). Since they are about between 2 to 4 times smaller than the above quoted neutral current pion-related initial backgrounds, very substantial background rejections of charged pions in the presence of the muon signal are required.

The precise determination of these remaining backgrounds requires very detailed physics emulations beyond the initial scope of the present paper. Therefore we shall limit ourselves here to some general considerations. In order to perform such a more detailed analysis actual neutrino events from a LAr-TPC detector are needed. The present ICARUS LAr-TPC experiment [22] at the LNGS has already collected a number of neutrino events (CC and NC) in which either muons and charged pions stop inside the fiducial volume. They should provide the necessary evidence for a better experimental proof of these backgrounds. Additional data may be provided in the near future with the ICARUS and MicroBooNe neutrino experiments located at a short distance neutrino beam and which may collect a much larger number of neutrino (CC and NC) events.

In view of the large mass of the LAr-TPC detector we may safely assume that all charged particles stop within the sensitive volume. Note that while the muon of the signal is a leading particle, the background pion is coming from target fragmentation. Therefore neutral current induced background events tend to pile up strongly in the region of small lepton energies, where the expected signal due to good $\nu_e \rightarrow \nu_\mu$ events is very small but also most sensitive to the specific physical alternatives.

In a LAr-TPC the energy deposited along the track and the length of the track are accurately measured with the help of a very large number of points, each about 3 mm apart, each corresponding to a LAr length of about 0.5 gr/cm$^2$. Heavier ioniza-tion losses have to be corrected by the known saturation effects in the track. Both $\pi^+$ and $\mu^+$ show the presence of a decay electron, although $\pi^+ \to \mu^+ + \nu$ is too short to be detectable in the LAr detector. Decay and nuclear interactions in flight of a pion can be easily identified and removed.

A first simple identification criterion is based on the fact that for a given observed kinetic energy, the residual range of a muon track is substantially longer than the one of a pion. In Figure 12 we show the distribution in the lengths of $\pi$ or $\mu$ tracks with a residual energy 100 MeV. The average lengths for a $\pi$ or $\mu$ are respectively 26 and 30 cm. Calculations using the FLUKA simulation [24] indicate that a separation is possible with a few percent loss. According to the calculation, for a range >28 gr/cm$^2$ about 95% of muons survive with a contamination of pions of $\approx$ 1/50 (90% c.l.). Higher rejections are of course possible with tighter muon cuts. The small, low range tail of the muon distribution overlapping with the pion line is due to the presence of the tail of energetic δ-rays, which should be generally directly identifiable: these events most likely could be removed.

A second criterion makes use of the fact that the LAr detector is segmented in a very large number of separate $dE/dx$ elements and a more sophisticated analysis, taking into account selectively of the individual Landau fluctuations provide selection criteria in addition to the above described simple range-energy sum. As well known the Bethe equation (BH) for the stopping power is a universal function of $(\beta\gamma)$ with the range proportional to the mass ($m_\pi/m_\mu = 1.32$). The experimental energy loss probability distribution of the LAr cells [22] is adequately described by the highly-skewed Landau (or Landau-Vavilov) distribution (Figure 13).

Electrons produced by an ionizing track may be partially recombined in the immediate vicinity of the creation point. Electron–ion recombination in the LAr-TPC chamber has been studied [26] both as a function of the electric field and as a function of the stopping power (dE/dx). Proton and muon tracks as well α particles from $^{241}$Am have been used to determine experimentally at E = 500 V/cm the recombination at different stopping powers segments were grouped and for each dE/dx bin the value for R was computed as the ratio between the mean measured and Bethe-Bloch formula estimated stopping power. The value of the experimental measurements and the values from the theoretical input Birks parameters are plotted as a function of the specific ionization loss [22] 14). The values are R =0.7 from minimum ionizing, decreasing to about R = 0.4 for dE/dx = 10 MeV/g/cm$^2$. These effects have been included in the calculations.

Selectivity to the mass of the particle is obtained with the help of a large number ($10^5$) Montecarlo generated $dE/dx_{actual,\mu}$ (namely BH + Landau-Vavilov fluctuations) distribution of muon events over the extended residual range $\leq 80$ gr/cm$^2$ (160 cells). In order to reduce the skew-ness due to the presence of energetic δ-rays, 20% of events in the tail have been removed. This distribution is then compared with the average value over the extended range of the Bethe equation (BH) predictions for the abscissa (Figure 15) assuming a muon $A_\mu = \left(dE/dx_{actual,\mu} - dE/dx_{BH,\mu}\right)\!\Big/dE/dx_{BH,\mu}$ or $B_\pi = \left(dE/dx_{actual,\mu} - dE/dx_{BH\pi}\right)\Big/dE/dx_{BH\pi}$ for a pion. While in the case of a muon the distribution is peaked to the value $\langle A_\mu \rangle \approx 0$, confirming the expectation, for a pion the distribution is shifted with respect to 0, namely $\langle B_\pi \rangle \approx 0.055$. According to the calculation, for ( $A_\mu \leq 0.018$ ) about 90% of muons survive with a contamination of pions of 1/320. For a tighter cut $A_\mu \leq 0.012$ the survival is 80% with a contamination of 1/1700.

Finally in the case of B-8 $\mu^-$ produced beam another additional criterion may be used. About 70% of the negative muons and all pions will undergo nuclear capture at the end of the range. While the $\pi^-$ related star involves the full pion rest mass into several heavy nuclear prongs, most of the $\mu^-$ related energy is emitted by the invisible neutrino. Discrimination of the local energy measured with LAr at the end point of the track may offer another powerful discrimination between $\pi^-$ and $\mu^-$. Optimisation of this method is still required: at present we assume provisionally at least an additional factor 50 in the $\pi^-$ misidentification.

We conclude that ample discrimination criteria are possible and that the discrimination between pions and muons should be amply sufficient in order to produce negligible backgrounds to the expected signal for the CP violating phase $\delta_{CP}$.

## 7.— Conclusions

An accelerator/storage setup based on β-decays of B-8 and Li-8 has been studied. The radioactive ions are produced in the "mirror" frame geometry, namely with a moving Lithium beam hitting a gas jet target. Kinematics is very favourable since the production angles of the secondary ions are emitted in a narrow angular cone and a low energy spectrum. They are absorbed as neutral atoms in a few tens of micron foils at high temperatures, from which they quickly escape to generate the ion source of the accelerator complex.

An accumulation storage ring pointing toward the distant neutrino detector has to be added to the accelerator complex. Decaying ions are accumulated stacking many successive pulses next to one another. The surviving fraction of ions is absorbed in a target dump. The main parameters of the acceleration and subsequent storage complex are shown in Table 1.

Rates have been estimated for an indicative accelerated ion intensity of $10^{13}$ ion/s and a far away LAr-TPC detector of 34 kton. In absence of sizeable backgrounds, the observations for the β-beam case should cover the full region of the oscillatory peaks relevant to the CP violating phase $\delta_{CP}$ with an adequate statistics and a reasonably sized detector. Event rates are generally at least one order of magnitude greater than the ones for a classic high power horn. The coverage in this latter case is much less complete and it essentially limited to the observations in the region around the first maximum.

The main signature relevant to the CP violating phase $\delta_{CP}$ is coming from $\nu_e \to \mu^- X$ from the B-8 and $\bar{\nu}_e \to \mu^+ X$ from the Li-8 initiated beam. The main competing background are the neutral current produced inelastic processes leading to a negative pion capture or positive pion decay, with the pion misidentified into a decaying or captured muon.

Several different methods can be used in order to separate out at the expected confidence level the pion contamination in the region of the track near to the end point, namely (1) the pion vs muon difference in range for a given deposited energy; (2) dE/dx Landau-Vavilov experimental distribution under the muon and pion hypothesis, averaging it over many ($\geq 100$) LAr cells and truncating the largest $\delta$-rays and (3) the differences in the visible energy in the capture stars for negative muons and pions.

The ICARUS LAr-TPC experiment [22] at the LNGS has recorded a large number of neutrino events (CC and NC) in which either muons or charged pions stop within the fiducial volume. A conclusive experimental analysis of these backgrounds will therefore be provided in the near future.

## Table 1.

| Beta beam source | B-8- | B-8 | B-8 | Li-8 | Li-8 | |
|---|---|---|---|---|---|---|
| Distance from beta source | 730 | 730 | 1290 | 730 | 1290 | km |
| Proton equivalent energy: | 120.00 | 190.00 | 190.00 | 400.00 | 400.00 | GeV |
| Accelerated ion intensity: | 1.0E13 | 1.0E13 | 1.0E13 | 1.0E13 | 1.0E13 | Ion/s |
| Gamma ion: | 79.940 | 126.57 | 126.57 | 159.88 | 159.88 | |
| Ion energy: | 600.00 | 950.00 | 950.00 | 1200.0 | 1200.0 | GeV |
| Bending radius: | 133.33 | 211.11 | 211.11 | 444.44 | 444.44 | m |
| Average bending dipole field | 3.0 | 3.0 | 3.0 | 3.0 | 3.0 | Tesla |
| Arc radius: | 166.67 | 263.89 | 263.89 | 555.56 | 555.56 | m |
| Bending circumference: | 1.0472 | 1.6581 | 1.6581 | 3.4907 | 3.4907 | km |
| Fraction neutrino in axis: | 0.333 | 0.333 | 0.333 | 0.333 | 0.333 | |
| SR Revolution time: | 10.472 | 16.581 | 16.581 | 34.907 | 34.907 | $\mu$s |
| Beam/s current to SR: | 763.94 | 482.49 | 482.49 | 137.51 | 137.51 | mA |
| Accelerated half-life: | 62.11 | 98.35 | 98.35 | 124.23 | 124.23 | sec |
| SR stored current: | 47.45 | 47.45 | 47.45 | 17.08 | 17.08 | A |
| Max. neutrino energy: | 2094.4 | 3316.2 | 3316.2 | 4188.9 | 4188.9 | MeV |
| Average neutrino energy: | 1047.2 | 1658.1 | 1658.1 | 2094.4 | 2094.4 | MeV |
| Angle for 1 cm at detector: | .137E-01 | .137E-01 | .775E-02 | .136E-01 | .775E-02 | $\mu$rad |
| Laboratory angle for 1 cm: | 2.1901 | 3.4677 | 1.9624 | 4.3803 | 2.4788 | $\mu$rad |
| Fraction neutrino in 1 cm2 | .127E-12 | .319E-12 | .103E-12 | .508E-12 | .163E-12 | |
| Neutrino flux: | 1.272 | 3.190 | 1.021 | 5.089 | 1.630 | n/sec/cm2 |
| Neutrino assumed CP angle: | 0.0 | 0.0 | 0.0 | 0.0 | 0.0 | degree |
| Neutrino flux: | 1.2724 | 3.1898 | 1.0215 | 5.0895 | 1.6298 | n/s/cm2/100MeV |
| Fiducial detector mass (LAr) | 34.0 | 34.0 | 34.0 | 34.0 | 34.0 | kton |
| Unoscillated event rate: | 24.930 | 92.025 | 29.470 | 74.545 | 23.872 | ev/day |
| $\nu_e \to \nu_\mu$ event rate: | 1.09 | 4.225 | 1.448 | 2.109 | 0.546 | ev/day |
| Unoscillated total: | 24930 | 92025 | 29470 | 74545 | 23872 | ev/5y/200d |
| **$\nu_e \to \nu_\mu$ total:** | **1088.7** | **4224.9** | **1447.8** | **2109.5** | **546.58** | **ev/5y/200d** |

# 8.— Figure captions.

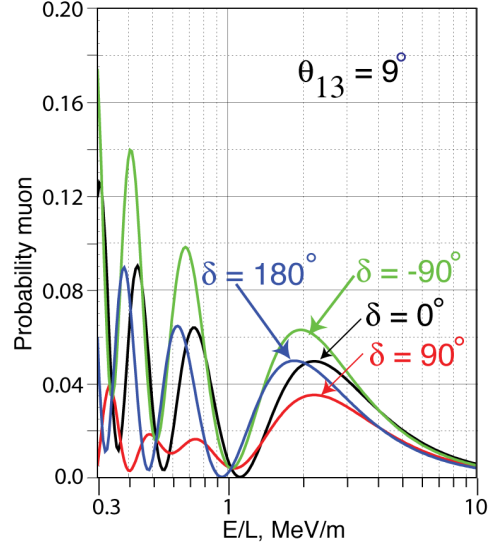

*Figure 1.* The CP violating amplitude $\delta$ for the measured [8] value $\sin^2(2\theta_{13}) = 0.0890 \pm 0.0112$ and with matter oscillation effects neglected. The oscillation mixing probability is given as a function of the distance in MeV/m for an initial $\nu_e$ conversion into a $\mu$ for absence of CP violation ($\delta = 0°$) and for three different values of the CP phase $\delta = 90°, -90°, 180°$.

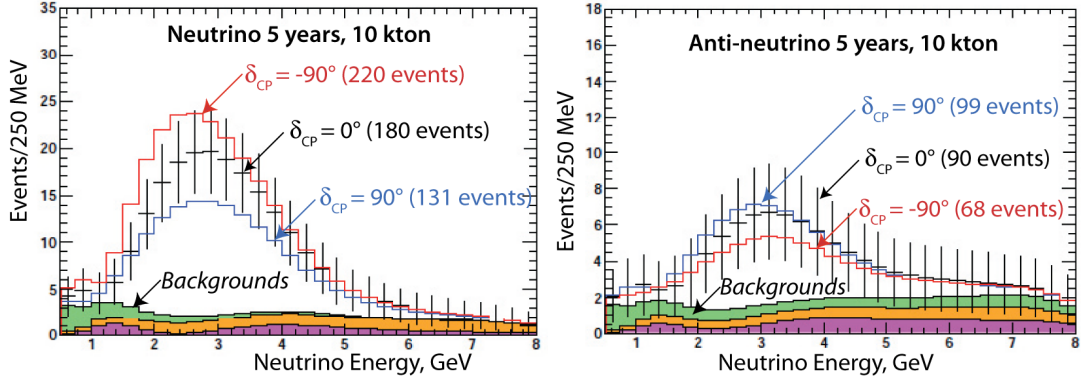

*Figure 2.* The LBNE [14] experiment at L=1290 km in Homestake and the beam from FNAL. Figures 2a and 2b show the expected spectra of $\nu_\mu \to \nu_e$ and $\bar{\nu}_\mu \to \bar{\nu}_e$ oscillation events in a 10-kton LArTPC for 5 years of neutrino and 5 years of antineutrino running with a 700 kW beam and normal mass ordering. Backgrounds from intrinsic beam $\nu_e$ (cyan), $\nu_\mu$-NC (yellow), and $\nu_\mu$ CC (green) are displayed as stacked histograms. In spite of the very long duration of the exposures, the event rates are rather small. The number of expected signal events for $E_\nu \leq 1.5$ GeV is statistically insignificant and it is dominated by backgrounds. The significance of a signal is therefore limited to the observation in the region with $E_\nu \geq 1.5$ GeV around the first maximum.

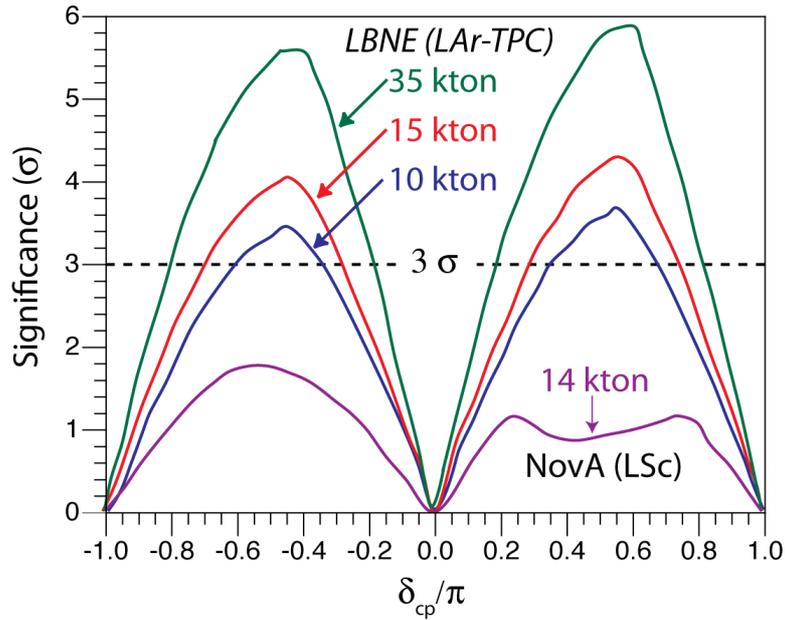

**Figure 3.-** The significance in terms of standard deviations of the predicted experimental result of LBNE experiment [14] as improved increasing the fiducial mass from 10 to 15 and 34 kton. The ability of determining $\delta_{CP}$ with appropriate statistical evidence is strongly dependent of the so far unknown value of $\delta_{CP}$. The expectations of the NOVA experiment [3] are also shown.

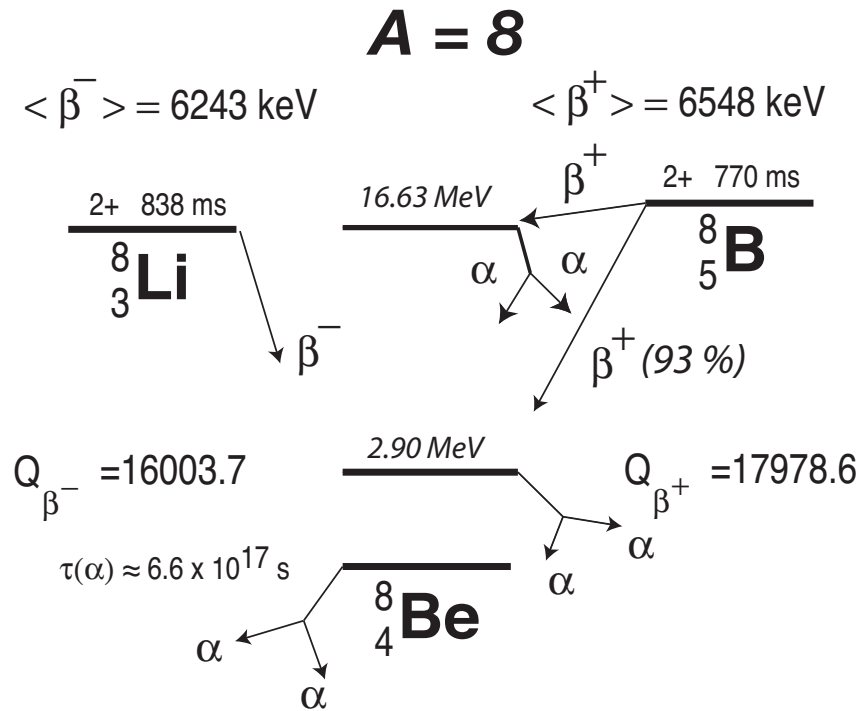

**Figure 4.-** Isospin triplet with A = 8 (Li-8, Be-8, B-8), decaying to the fundamental level of Be-8. In absence of Coulomb corrections, the three states would have identical nucleons configurations because of charge independence. The actual experimental values of the beta decaying doublet Li-8 with $\tau_{1/2}$ = 0.84 s and B-8 with $\tau_{1/2}$ = 0.77 s are respectively $Q^*$ = 16.005 MeV and $Q^*$ = 16.957 MeV.

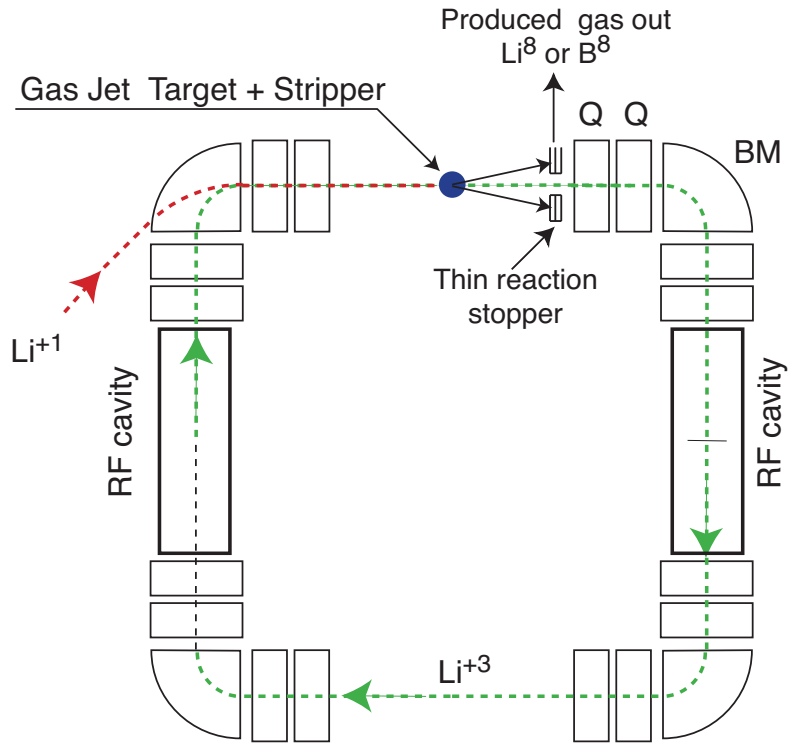

***Figure 5.*** Principle diagram [17] of the ionization cooled ion accumulation in the small FODO storage ring. Singly charged ions from a low energy accelerator are fully stripped by a thin gas jet target and stably stored inside the storage ring. The energy is continuously lost by the gas target and recovered by a RF cavity, until a nuclear collision ejects the particle from the gas. Radioactive ions produced in the gas jet target are accumulated in the thin reaction stopper and quickly escape from the thin foil as neutral gas. The recovered gas will be later ionized again and accelerated to high energies. This technique of using very thin targets in order to produce secondary neutral beams has been in use for many years by the ISOLDE team.

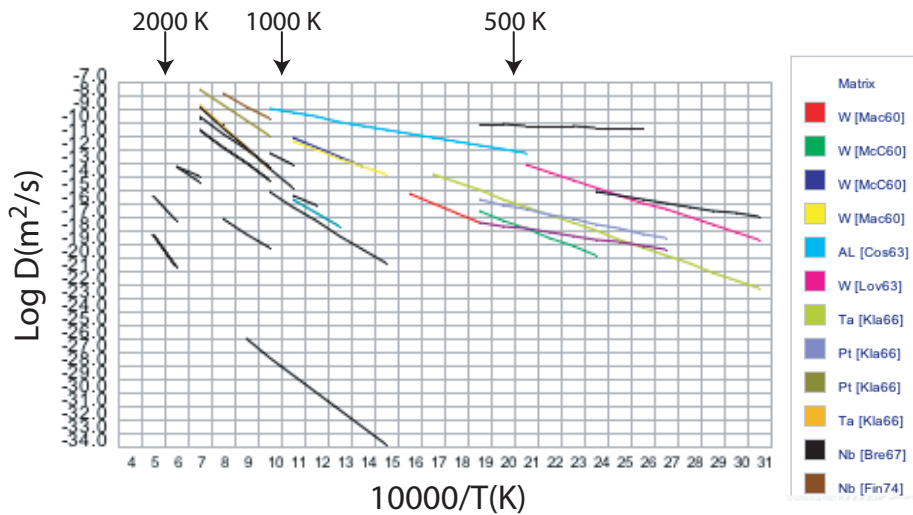

***Figure 6.-*** Fink diffusion coefficient D (m$^2$/s) for Lithium and different materials.[19]

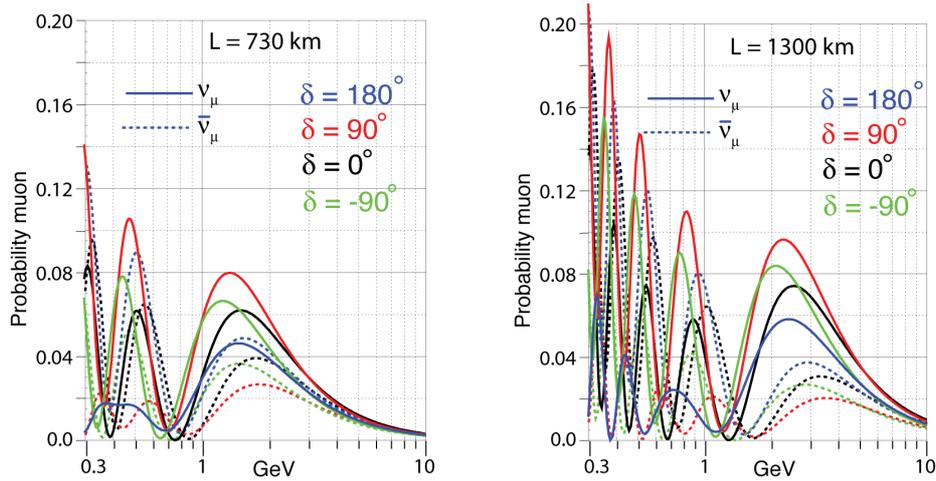

***Figure 7.-*** Two indicative distances L of the detector from the neutrino source have been considered, namely LNGS/CERN or FNAL/Soudan with L = 730 km in Figure 7a and FNAL/Homestake with L = 1290 km in Figure 7b. The neutrino (continuous) and antineutrino (dotted) spectra for the process are shown in absence of CP violation) and for three different values of the CP violating phase δ = 90°, -90°,180°, as a function of the energy in GeV.

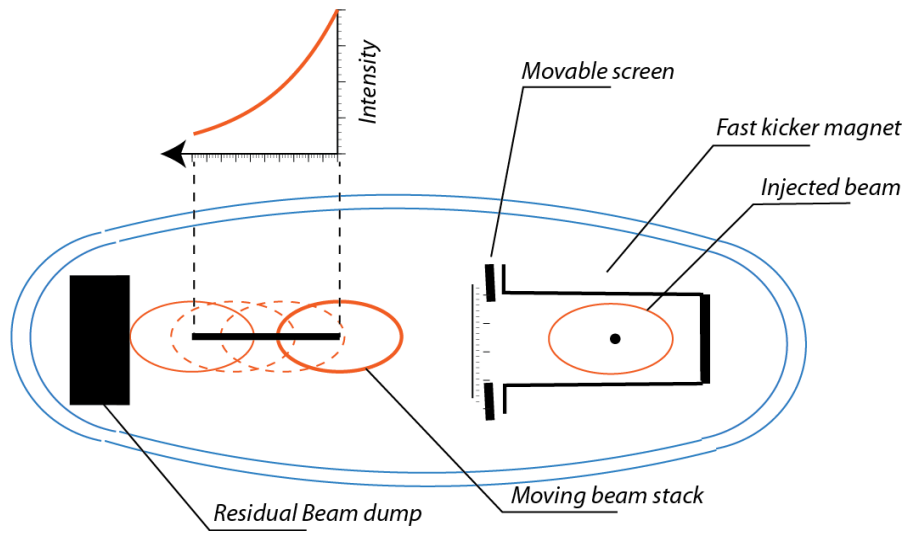

***Figure 8.-*** The transfer from the accelerator to the storage ring (SR). The circulating currents in the SR are similar to the one of the protons of the CERN-ISR [20] but with a much larger and superconducting ring. After a pulse has been injected from the accelerator, a RF system accelerates the protons just enough to move the particles with a moving septum from their injection orbit to an orbit nearer to the outside of the vacuum chamber. The injection orbit is then ready to receive the next accelerator pulse, in turn accelerated and moved to an orbit only a fraction of mm from where the previous pulse was left. Since ions β-decay with a lifetime of the order of 60 to 120 s depending on the configuration, the stacking process can be repeated indefinitely. The surviving ions are removed with an internal beam dump oriented toward the direction opposite from the one of the decay of the neutrino beam.

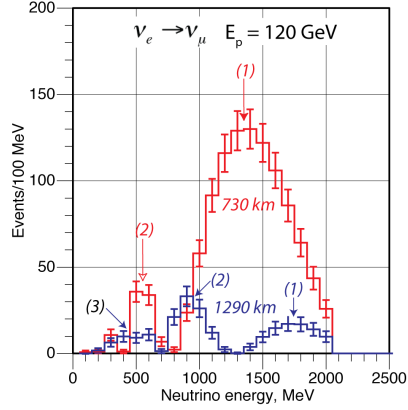

***Figure 9.-*** Differential oscillated spectra (events/100 MeV) as a function of the B-8 neutrino energy detected 730 km and 1290 km away from the 120 GeV equivalent protons of the FNAL Main Energy Injector. Events rates taken from Table 1 are given for 5 years of exposure, 200 days/year in a 34 kton fiducial LAr-TPC detector and $\theta_{13}$= 9°.

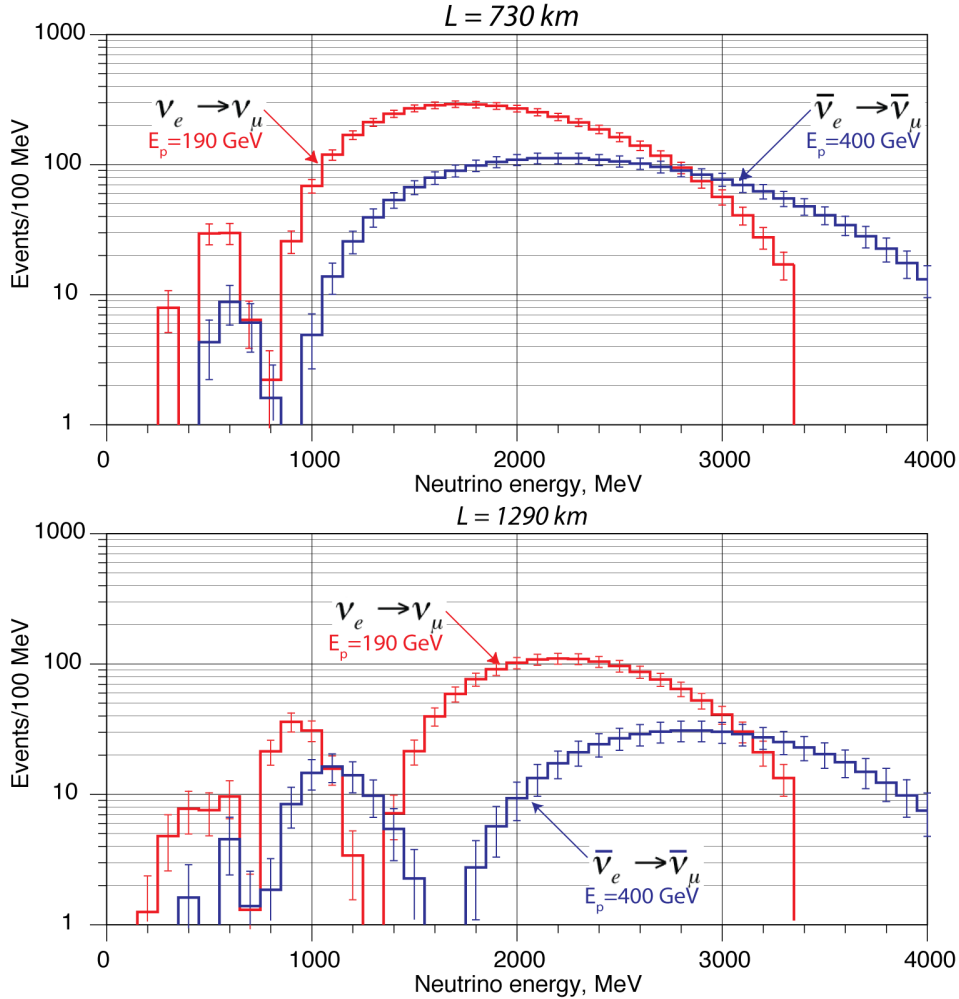

***Figure 10.-*** Differential oscillated spectra (events/100 MeV) detected 730 km (Figure 10a) and 1290 km (Figure 10b) away from the CERN accelerator complex. Two equivalent proton energies have been considered: for B-8 (neutrino) we have chosen Ep =190 GeV; for Li-8 (anti-neutrino)we have chosen Ep= 400 GeV. Events rates taken from Table 1 are given for 5 years of exposure, 200 days/year in a 34 kton fiducial LAr-TPC and $\theta_{13}$ = 9°.

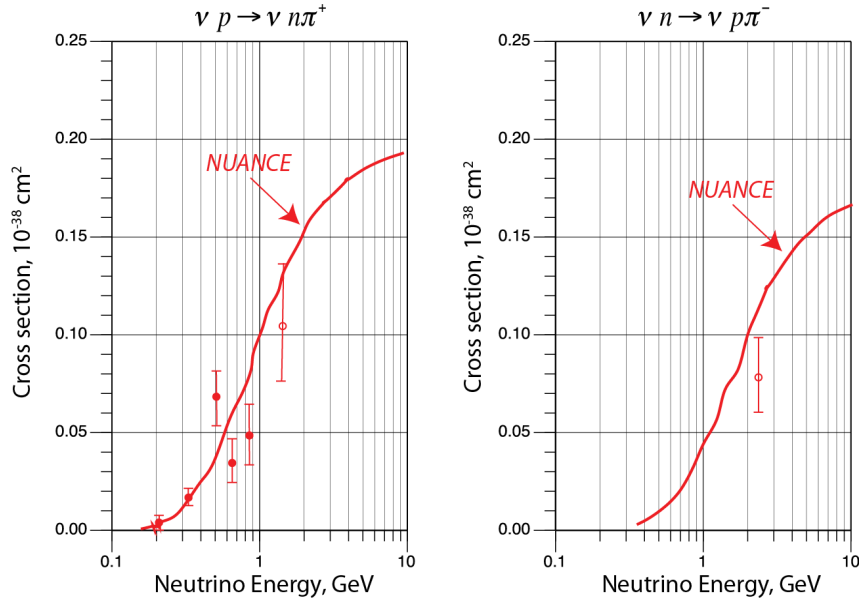

***Figure 11.-*** Single charged pion production from neutral current events potentially simulating the muon track of an oscillated transition. Actual measurements are from Gargamelle [23] (open circles) and from ANL [23] closed circles. Simulations have been generated with NUANCE [24].

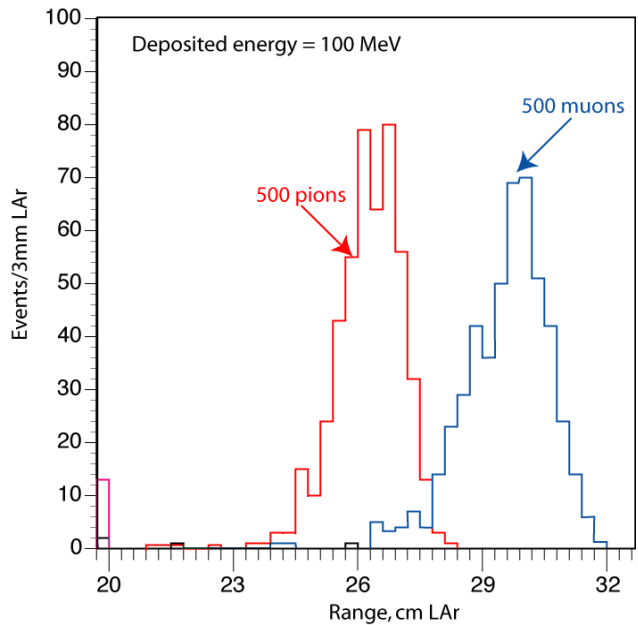

***Figure 12.-*** Range distribution in cm of LAr-TPC for pions (red) and muons (blue) and 100 MeV of deposited energy. The calculations take correctly into account all relevant phenomena of energy losses with the help of a FLUKA Monte Carlo programme [25]. It is apparent that pions and muons have clearly distinguishable ranges. According to the calculation, for a range >28 gr/cm² about 95% of muons survive with a contamination of pions of ≈ 1/50 (90% c.l.). Higher rejections are of course possible with tighter muon cuts. The small, low range tail of the muon distribution overlapping with the pion line is due to the presence of the tail of energetic δ-rays, which are generally directly identifiable and these events most likely could be removed.

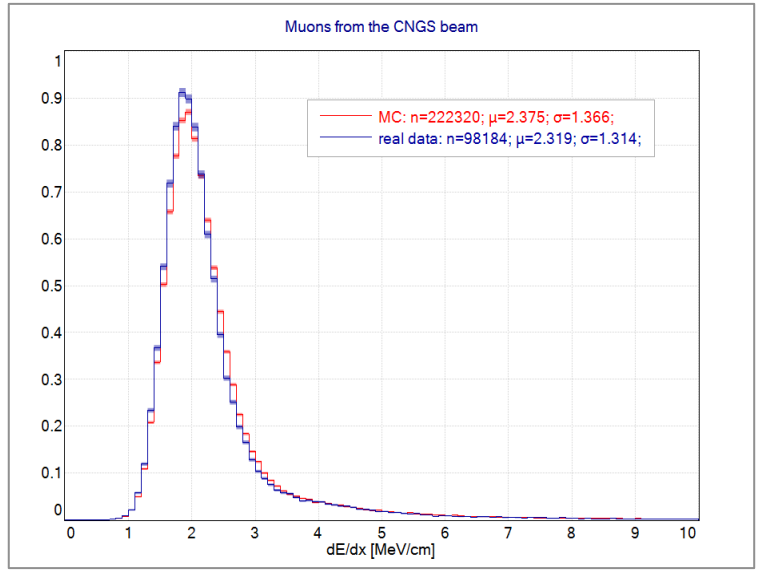

***Figure 13.-*** Experimental distribution of individual cells of relativistic muons from the ICARUS LAr-TPC experiment [22] at the LNGS (blue) compared with the theoretical predictions coming from the Landau-Vavilov distribution (red). The two distributions are in excellent agreement.

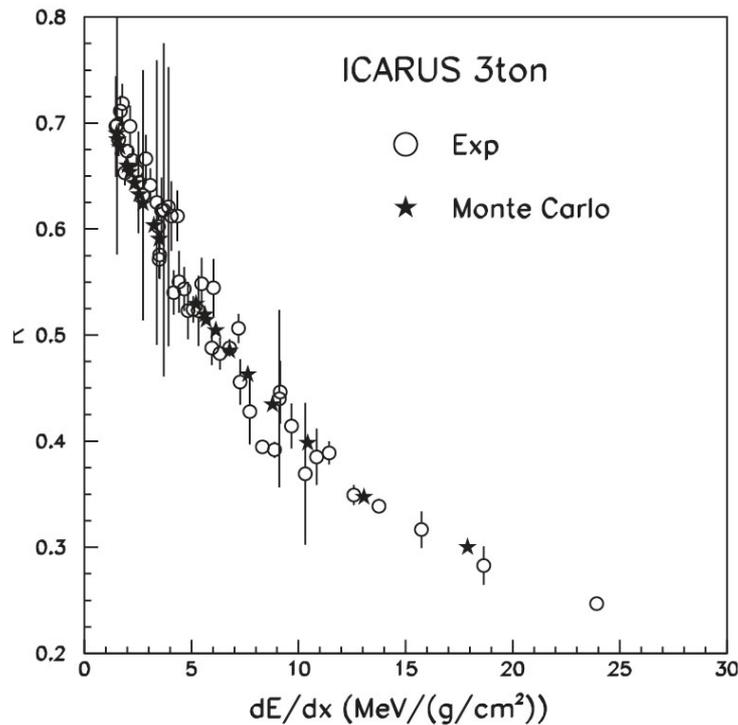

***Figure 14.*** Electron–ion recombination in the LAr-TPC chamber has been studied [26] both as a function of the electric field and as a function of the stopping power (dE/dx). Proton and muon tracks have been used to determine experimentally in the range 0.1 to 1.0 kV/cm and dE/dx from 1.5 to 30 MeV/g cm$^2$. Data have been fitted to a Birks law as a function of the particle stopping power and of the drift electric field. The agreement between data points and the fitted functions for a field of 500 V/cm is remarkably good [26].

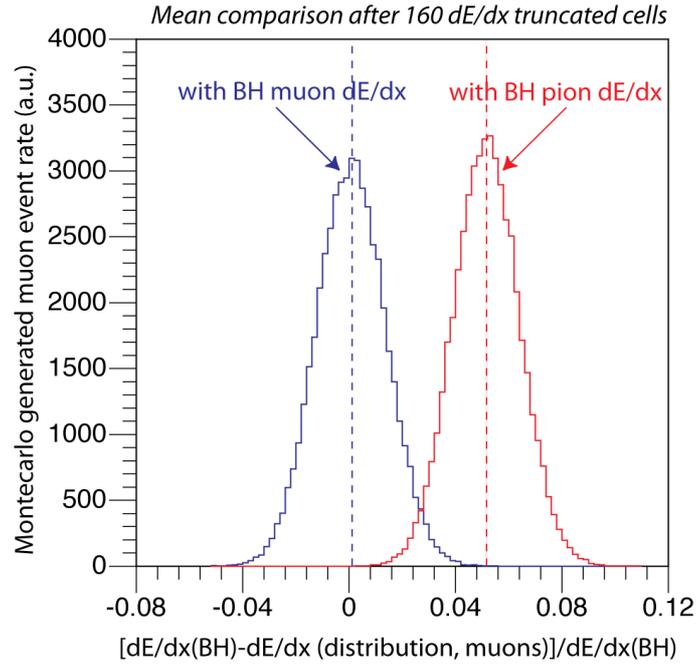

***Figure 15.*** Selectivity to the mass of the particle obtained with the help of a large number ($10^5$) Montecarlo generated dE/dx (namely BH + Landau-Vavilov fluctuations) event distribution of muons over an extended residual range, typically $\leq 80$ gr/cm$^2$ (160 cells). In order to reduce the skew-ness due to the presence of energetic δ-rays, 20% of events with the largest dE/dx in the tail have been removed. The resulting muon distribution is compared with the predictions of the Bethe equation (BH), chosing as a value for the abscissa either a muon or a pion. While in the case of a muon the distribution is peaked to the value $<A_\mu> \approx 0$, confirming the expectations, for a pion the distribution is shifted with respect to 0, namely $<B_\pi> = 0.055$. According to the calculation, for $A_\mu \leq 0.018$ about 90% of muons survive with a contamination of pions of 1/320. For a tighter cut $A_\mu \leq 0.012$ the survival is 80% with a contamination of 1/1700.